\documentclass[aps,twocolumn,showpacs,preprintnumbers,floats]{revtex4}
\usepackage{amssymb}
\usepackage{graphicx}
\usepackage{dcolumn}
\usepackage{bm,CJK}
\usepackage{epsfig}
\usepackage{hyperref}
\usepackage{graphicx}
\usepackage{dcolumn}
\usepackage{bm}
\usepackage{longtable}
\usepackage[dvips]{color}
\usepackage[normalem]{ulem}

\newcommand{\PLB}[3]{Phys.\ Lett.\ B\ {\bf #1}, #2 (#3)}
\newcommand{\PRT}[3]{Phys.\ Rep.\ {\bf #1}, #2 (#3)}
\newcommand{\PRL}[3]{Phys.\ Rev.\ Lett.\ {\bf #1}, #2 (#3)}

\newcommand{\PRC}[3]{Phys.\ Rev.\ C\ {\bf #1}, #2 (#3)}
\newcommand{\PRD}[3]{ Phys.\ Rev.\ D\ {\bf #1}, #2 (#3)}
\newcommand{\JPG}[3]{J.\ Phys.\ G\ {\bf #1}, #2 (#3)}

\newcommand{\JHEP}[3]{J.\ High\ Energy\ Phys.\ {\bf #1} (#3)\ #2}

\begin{document}

\title{Effects of the temperature and magnetic-field dependent coupling on the properties of QCD matter}
\date{\today}
\author{Li Yang,
Xin-Jian Wen\footnote{wenxj@sxu.edu.cn} }
\affiliation{Institute of
Theoretical Physics, Shanxi University, Taiyuan 030006, China }

\begin{abstract}
To reflect the asymptotic freedom in the thermal direction, a
temperature-dependent coupling was proposed in the literature. We
investigate its effect on QCD matter with and without strong
magnetic fields. Compared with the fixed coupling constant, the
running coupling leads to a drastic change in the dynamical quark
mass, entropy density, sound velocity, and specific heat. The
crossover transition of QCD matter at finite temperature is
characterized by the pseudocritical temperature $T_\mathrm{pc}$,
which is generally determined by the peak of the derivative of the
quark condensate with respect to the temperature $d\phi/dT$, or
equivalently, by the derivative of the quark dynamical mass $d
M/dT$. In a strong magnetic field, the temperature- and
magnetic-field-dependent coupling $G(eB,T)$ was recently introduced
to account for inverse magnetic catalysis. We propose an analytical
relation between the two criteria $d\phi/dT$ and $dM/dT$ and show a
discrepancy between them in finding the pseudocritical temperature.
The magnitude of the discrepancy depends on the behavior of $dG/dT$.
\end{abstract}

\pacs{12.39.-x, 12.40.jn, 12.38.Mh}
\maketitle

%\textbf{Key words:} pseudocritical temperature, strong magnetic field, Nambu$%
%-$Jona-Lasinio model

\section{INTRODUCTION}
It is well known that with increasing baryon number density and
temperature, hadronic matter undergoes a phase transition to
quark-gluon plasma. In high temperature or high densities regions,
the asymptotic freedom becomes important in the investigation of the
QCD diagram \cite{Veronique}. To obtain a comprehensive QCD diagram,
it is necessary to understand the phase transition in the presence
of magnetic fields. In recent years, research related to strong
magnetic fields has been carried out in both condensed matter
physics \cite{Lai01} and the particle physics \cite{Miransky15}.
Strong magnetic fields could have a drastic influence on the special
stability of quark matter \cite{bord11,Felipe09,wen13}, the
anisotropy of the equation of state \cite{Meneze15}, the region of
the phase transition, and (inverse)magnetic catalysis. The presence
of magnetic fields can promote a change in the size and location of
the first-order line \cite{Ferra12} and increase the mass of neutron
stars and white dwarf stars beyond the Chandrasekhar's limit
\cite{das12}. Both thermodynamical and dynamical quantities display
an oscillating behavior in the presence of magnetic fields
\cite{Ebert99}. Magnetic catalysis was found  to have an important
effect on chiral symmetry breaking enhanced by an external magnetic
field \cite{Miransky94,Miransky,Inagaki,catapaper}.

The Nambu$-$Jona-Lasinio(NJL) model has been successful in
investigating the QCD diagram, and recently it was extended to
easily reproduce the behavior of the quark condensate and the
dynamical mass with an external magnetic field
\cite{Meneze09,allen13}. It was further extended by including tensor
channels (which leads to a spin-one condensate \cite{Ferrer}) or by
including the eight-quark interaction \cite{Gatto}. In addition to
the magnetic effect at vanishing chemical potential, inverse
magnetic catalysis was initially suggested as a mechanism to
decrease the critical chemical potential for chiral restoration
\cite{Preis}. It was later predicted by lattice simulations at zero
density that the critical temperature of the chiral transition
decreases with the magnetic field \cite{Bali}. It has also recently
attracted much theoretical attention in various phenomenological
models \cite{wen2015,pagu17}. It is apparent that the failure of the
previous effective models to provide inverse magnetic catalysis can
be attributed to the fact that the coupling constant does not run
with the magnetic field \cite{Farias}, or strictly speaking, the
effective models lack gluonic degrees of freedom and can not account
for the backreaction of sea quarks to the magnetic field
\cite{Farias16}.

Many attempts have been make to interpret the inverse magnetic
catalysis. One approach is through a magnetic-field- and
temperature- dependent coupling \cite{Farias, Farias16} or a
parametrized fitting function \cite{Ferreira} taking into account
the asymptotic-freedom effect near the critical point. Another
approach is through the parametrization of the Polyakov loop ( whose
coefficients depend on both the temperature and magnetic field) to
mimic the reaction of the gluon sector to the magnetic field
\cite{Ferre14}. In fact, the employment of asymptotic freedom in the
phenomenological approach can be traced to early works in the
literature. For example, the QCD coupling was introduced to depend
on the environmental parameters, such as the density \cite{bann08},
temperature \cite{enqv92}, and magnetic field
\cite{Miransky,Ferreira}. Based on the general argument from the
renormalization group equation \cite{Coll75}, these characteristics
replace the momentum as the running scale. The special running
behavior will lead to a detailed change in the properties of QCD
matter. The magnetic-field-dependent running coupling reveals
interesting properties, such as a change in the dynamical mass and
the stability of magnetized quark matter
\cite{wen2016a,wen2016b,yang}. One recent work reported that the
magnetization changes due to the variation of the coupling constant
with respect to the magnetic field $\partial G/\partial B$
\cite{Farias16}. Then, one may ask about the contribution of
$\partial G/\partial T$. In this paper, we first analyze the
behavior of the coupling dependent on the temperature in two-flavor
quark matter. Then, we investigate the influence of the temperature-
and magnetic-field-dependent coupling on the pseudocritical
temperature of the crossover transition in a strong magnetic field.

This work is organized as follows. In Sec. \ref{sec2}, We briefly
review the NJL model of quark matter in both zero magnetic field and
a strong magnetic field. Correspondingly, the two kinds of running
couplings are introduced as well as the model parameters in the
computation. In Sec. \ref{sec:result}, the numerical results and a
discussion are given, with a detailed analysis of the effects of the
running coupling on the thermodynamical quantities. The last section
is a short summary.

\section{Thermodynamics of the SU(2) NJL Model}
\label{sec2}
\subsection{Thermodynamics of the SU(2) NJL Model in zero magnetic field}
In the SU(2) version of the NJL model without a magnetic field, the
Lagrangian density of the two-flavor NJL model is given by
\begin{equation}
{\mathcal{L}}_{NJL}=\bar{\psi}(i/\kern-0.5em\partial -m)\psi +G[(\bar{\psi}%
\psi )^{2}+(\bar{\psi}i\gamma _{5}\vec{\tau}\psi )^{2}].
\end{equation}%
where $\psi $ represents a flavor isodoublet ($u$ and $d$ quarks) and $%
\vec{\tau}$ are isospin Pauli matrices. In the mean-field
approximation \cite{Ratti}, the dynamical quark mass is
\begin{equation}
M_i=m-2G\langle \bar{\psi}\psi \rangle .  \label{eq:gap}
\end{equation}
where the quark condensates include $u$ and $d$ quark contributions
as $\langle \bar{\psi}\psi \rangle \equiv\phi=\sum_{i=u,d}\phi
_{i}$. The dynamical mass depends on both flavor condensates.
Therefore, the same mass $ M_{u}=M_{d}=M$ is available for $u$ and
$d$ quarks. The contribution from the quark with flavor $i$ is
\begin{equation}
\phi _{i}=\phi _{i}^{\mathrm{vac}}+\phi _{i}^{%
\mathrm{med}}.  \label{eq:condensate}
\end{equation}
The terms $\phi_i ^{\mathrm{vac}}$ and $\phi_i ^{%
\mathrm{med}}$ represent the vacuum and medium contributions to the
quark condensation, respectively,
\begin{eqnarray}
\phi_i ^{\mathrm{vac}} =-\frac{MN_{c}}{2\pi ^{2}}\left[ \Lambda \sqrt{%
\Lambda ^{2}+M^{2}}-M^{2}\ln (\frac{\Lambda +\sqrt{\Lambda ^{2}+M^{2}}}{M})%
\right], \label{eq:convac}\nonumber
\hspace{-1.47cm}\\
\end{eqnarray}
\begin{eqnarray}
\phi_i ^{\mathrm{med}} =\frac{2MN_{c}}{\pi ^{2}}\int_{0}^{\infty }\frac{f}{%
E^{\ast }}p^{2}dp.
\end{eqnarray}%
where the effective quantity is $E^{\ast }=\sqrt{p^{2}+M^{2}}$, and
the fermion distribution function is defined as
\begin{equation}
f=\frac{1}{1+\exp [E^{\ast }/T]}.
\end{equation}

The total thermodynamic potential density in the mean-field
approximation reads
\begin{equation}
\Omega =\frac{(M-m_{0})^{2}}{4G}+\sum_{i=u,d}\Omega _{i},
\label{omega}
\end{equation}%
where the first term is the interaction term. In the second term,
$\Omega_i$ is defined as $\Omega _{i}=\Omega
_{i}^{\mathrm{vac}}+\Omega _{i}^{\mathrm{med}}$. The vacuum and
medium contributions to the thermodynamic potential are
\begin{eqnarray}
\Omega_i ^{\mathrm{vac}}&=&\frac{N_{c}}{8\pi ^{2}}\left[ M^{4}\ln
(\frac{\Lambda +\epsilon _{\Lambda }}{M})-\epsilon _{\Lambda
}\Lambda (\Lambda ^{2}+\epsilon _{\Lambda }^{2})\right] ,\\
\Omega_i ^{\mathrm{med}}&=&-\frac{2TN_{c}}{\pi ^{2}}\int_{0}^{\infty
}\left\{ \ln \Big[1+\exp (-\frac{E^{\ast }}{T})\Big]\right\}
p^{2}dp.\nonumber\\
\end{eqnarray}%
where the quantity $\epsilon _{\Lambda }$ is defined as $\epsilon
_{\Lambda}=\sqrt{\Lambda ^{2}+M^{2}}$. The ultraviolet divergence in the vacuum part $%
\Omega_i ^{\mathrm{vac}}$ of the thermodynamic potential is
removed by the
momentum cutoff. The effective pressure in the system is corrected by defining $P^{\mathrm{eff}%
}(T)=P(T)-P(0)$. The sound velocity, specific heat, and entropy
density from the flavor $i$ contribution are given as
\cite{Hatsuda,Farias16}
\begin{eqnarray}
c_s^2 &=&  \frac{\partial P^\mathrm{eff}}{\partial \epsilon}\Big|_S, \ \ \ \ C_V=T\frac{\partial^2 P^\mathrm{eff}}{\partial T^2}\Big|_V,\\
 S_i &=&
-\frac{2N_c}{\pi^{2}}\int_{0}^{\infty}\left[f\ln(f)+(1-f)\ln(1-f)\right]p^{2}dp.\nonumber\\
\end{eqnarray}

In principle, the interaction coupling constant between quarks
should be solved by the renormalization group equation, or it can be
phenomenologically expressed in an effective potential dependent on
environmental variables \cite{Richard,sinha,Xujf}. In the infrared
region, the nonperturbative effect becomes important and the
dynamical gluon mass represents the confinement feature of QCD
\cite{Natale}. Here we adopt the temperature-dependent running
coupling to investigate the thermal effect in the high-temperature
region \cite{Veronique},
\begin{equation}
G^{\prime }(T)={G_0}\sqrt{1-(\frac{T}{T_0})^{2}},
\end{equation}
where $T_0=0.3\Lambda$ is the critical temperature.

\subsection{Thermodynamics of the SU(2) NJL model in a strong magnetic field}

In the presence of strong external magnetic fields, the Lagrangian
density of the two-flavor NJL model in a strong magnetic field is
given as
\begin{equation}
{\mathcal{L}}_{NJL}=\bar{\psi}(i/\kern-0.5emD-m)\psi +G[(\bar{\psi}\psi
)^{2}+(\bar{\psi}i\gamma _{5}\vec{\tau}\psi )^{2}].
\end{equation}%
where the covariant derivative $D_{\mu }=\partial _{\mu
}-iq_{\mathrm{i}}A_{\mu }$ represents the coupling of the quarks to
the electromagnetic field, a sum over flavor and color degrees of
freedom is implicit. The dynamical quark mass is the same as
Eq.~(\ref{eq:gap}), but the quark condensates should include an
additional term from the magnetic field contribution,
\begin{equation}
\phi _{i}=\phi _{i}^{\mathrm{vac}}+\phi _{i}^{\mathrm{mag}}+\phi _{i}^{%
\mathrm{med}},  \label{eq:condensate}
\end{equation}
where the vacuum contribution $\phi _{i}^{\mathrm{vac}}$ is the same
as Eq.~(\ref{eq:convac}). The magnetic field and medium
contributions to the quark condensation are \cite
{Meneze09,menezes11}
\begin{eqnarray}
\phi _{i}^{\mathrm{mag}} &=&-\frac{M|q_{i}|BN_{c}}{2\pi ^{2}}\left\{
\ln [\Gamma (x_{i})]-\frac{1}{2}\ln (2\pi)+x_{i}\right. \nonumber\\
 & & \left. -\frac{1}{2}(2x_{i}-1)\ln
(x_{i})\right\} , \\
\phi _{i}^{\mathrm{med}} &=&\sum_{k_{i}=0}a_{k_{i}}\frac{M|q_{i}|BN_{c}}{%
2\pi ^{2}}\int \frac{f_i}{E_{i}^\ast} dp.
\end{eqnarray}
where $a_{k_{i}}=2-\delta_{k0}$ and $k_{i}$ are the degeneracy label
and the Landau quantum number, respectively. The dimensionless
quantity $x_i$ is defined as $x_i=M^{2}/(2|q_{i}|B)$. It can be seen
that the quark condensation is greatly strengthened by the factor
$|q_{i}B|$ together with the dimensional reduction $D-2$
\cite{Miransky,Kojo14}. In the second equation above, the
temperature contribution with zero chemical potential is introduced
in the fermion distribution function as
\begin{eqnarray}
f_i=\frac{1}{1+\exp[E_i^\ast/T]}.
\end{eqnarray}
The effective quantity $E_i^\ast=\sqrt{p^{2}+s_{i}^{2}}$ sensitively
depends on the magnetic field through
$s_{i}=\sqrt{M^{2}+2k_{i}|q_{i}|B}$.

Accordingly, the thermodynamic potential density $\Omega_i$ becomes
a sum of three terms,
\begin{equation}
\Omega _{i}=\Omega _{i}^{\mathrm{vac}}+\Omega _{i}^{
\mathrm{mag}}+\Omega _{i}^{\mathrm{med}}.
\end{equation}
where only the second and third terms feel the strong magnetic field
and should be rewritten as
\begin{widetext}
\begin{eqnarray}
\Omega _{i}^{\mathrm{mag}}&=&-\frac{N_{c}(|q_{i}|B)^{2}}{2\pi
^{2}}\left[
\zeta^{\prime}(-1,x_i)-\frac{1}{2}(x_{i}^{2}-x_{i})\ln (x_{i})+\frac{%
x_{i}^{2}}{4}\right],\\
 \Omega _{i}^{\mathrm{med}}
&=&-T\sum_{k=0}a_{k_{i}}\frac{|q_{i}|BN_{c}}{2\pi ^{2}}\int
dp\left\{ \ln \Big[1+\exp (-\frac{E_{i}^{\ast }}{T})\Big]\right\}.
\end{eqnarray}
\end{widetext}
where $\zeta (a,x)=\sum_{n=0}^{\infty }\frac{1}{(a+n)^{x}}$ is the Hurwitz
zeta function.

In the presence of a strong magnetic field, it is well known that
the interaction constant shows an obvious decreasing behavior in
addition to the enlargement of the gluon mass \cite{wen2015}. For
sufficiently strong magnetic fields $eB\gg \Lambda
_{\mathrm{QCD}}^{2}$, it is reasonable to express the coupling
constant $\alpha _{s}$ related to the magnetic field \cite{Miransky,Ferreira}%
. Motivated by the work of Miransky and Shovkovy \cite{Miransky}, a
similar ansatz for the magnetic-field-dependent coupling constant
was introduced in the SU(2)NJL models \cite{Farias16}:
\begin{equation}
G(eB,T)=c(B)\left[ 1-\frac{1}{1+e^{\beta (B)[T_{a}(B)-T]}}\right]
+s(B),\label{eq:Grun}
\end{equation}%
where the four parameters $c$, $\beta$, $T_a$, and $s$ were obtained
by fitting the lattice data and are strongly dependent on the magnetic field \cite%
{Farias16}.

To identify the pseudocritical temperature of the crossover transition, one
generally uses the location of the peaks for the vacuum quark condensates $%
|\langle \bar{\psi}\psi \rangle|$ \cite{Gatto}, or the normalized quark
condensates \cite{Gatto11,Ferre14},
\begin{eqnarray}
\sigma=\frac{\langle \bar{\psi}\psi \rangle(B,T)}{\langle \bar{\psi}\psi
\rangle(B,0)},
\end{eqnarray}%
which means that the quark condensate is measured in units of the
condensate at $T=0$. In fact, the crossover is signaled by a rapid
increase of the energy density. Thus, it has been suggested that the
crossover transition is determined by the maximum of $-dM/dT$
\cite{Ferra12}, which is generally consistent with $d\phi/dT$
\cite{avanci12}. However, when the coupling constant runs with the
temperature, a discrepancy will appear between them. From Eq.\
(\ref{eq:gap}), we obtain the following relation:
\begin{eqnarray}
\frac{d M}{dT}=-2G\frac{d \langle \bar{\psi}\psi \rangle}{dT}-2 \langle \bar{%
\psi}\psi \rangle \frac{dG}{dT},  \label{eq:new}
\end{eqnarray}
where the additional second term is necessarily introduced by the
temperature dependence of the running coupling, and will lead to a new
formula for the determination of the pseudocritical temperature in the next
section.

\section{Numerical Results and Discussion}
\label{sec:result}

For the SU(2) NJL model in this paper, we adopt the parameters $\Lambda=650 $%
MeV , $m_u=m_d=5.5$MeV, and $G_0=4.50373$ GeV$^{-2}$ in the
calculation. In order to reflect the asymptotic freedom in the
thermal region, the two kinds of running couplings are adopted for
the zero magnetic field case and strong magnetic field case
\cite{Veronique,Farias16}. The temperature dependence of
$G^{\prime}(T)$ and the thermomagnetic dependence of $G(eB,T)$ were
obtained by fitting lattice QCD predictions for the chiral
transition order parameter.

\subsection{ In zero magnetic field}
\begin{figure}[tbp]
\begin{center}
\includegraphics[width=0.40 \textwidth]{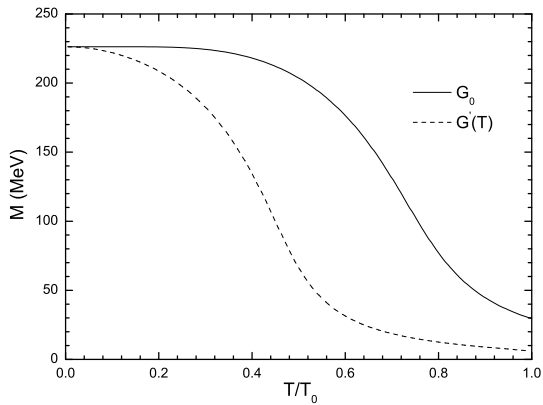}
\end{center}
\caption{{\protect\footnotesize The dynamical mass of the quark
versus temperature for the coupling constant $G_0$ and the running
coupling $G'(T)$. The critical temperature $T_0$ is 195 MeV.}}
\label{Fig1}
\end{figure}

The quark condensate or the dynamical mass is usually considered as
an order parameter of the chiral phase transition. The dynamical
mass decreases as the temperature increases, and the
chiral-symmetric phase is restored. In this section, we mainly
discuss the chiral restoration under the coupling constant $G_0$ and
the temperature-dependent running coupling $G^\prime(T)$ in zero
magnetic field. The dynamical quark mass $M$ is shown as a function
of the temperature in Fig.~\ref{Fig1}. The solid and dashed lines
are for the fixed coupling constant $G_0$ and the running coupling
$G'(T)$, respectively. It is clear that the absolute value of the
quark condensate under the running coupling $G'(T)$ is lower, which
increases the possibility pf having massless mass compared to the
fixed coupling constant $G_0$ case. Thus, the chiral-restoring
transition can be realized ata lower temperature with the running
coupling in our considerations.

\begin{figure}[htp]
\begin{center}
\includegraphics[width=0.40 \textwidth]{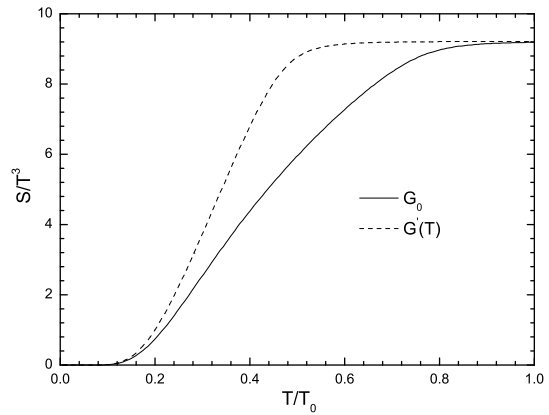}
\end{center}
\caption{{\protect\footnotesize The entropy density divided by $T^3$
as a function of the temperature for the couplings $G_0$ and
$G'(T)$.}} \label{Fig2}
\end{figure}

According to the second law of thermodynamics, entropy is an
increasing dimensionless function of temperature. In our work, we
use the ratio of the entropy density and the cube of the temperature
to get a dimensionless quantity in Fig. \ref{Fig2}. The ratio
increases and reaches a constant value ($S/T^3=9.2$) as the
temperature increases. However, in the temperature range of $80 \sim
120$ MeV, the dashed line for the running coupling $G'(T)$ is higher
than solid line for the fixed coupling $G_0$ case. The entropy
density is increased by the running coupling $G'(T)$, which can be
understood from the fact that the temperature-dependent interaction
strength becomes weak enough as the temperature increases.

\begin{figure}[bhtp]
\begin{center}
\includegraphics[width=0.44 \textwidth]{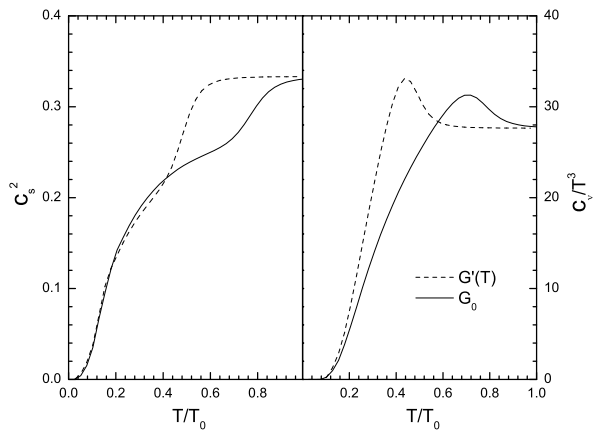}
\end{center}
\caption{{\protect\footnotesize Sound velocity and specific heat
versus temperature. The two lines are coincident at $T_0$.}}
\label{Fig3}
\end{figure}

In Fig.~\ref{Fig3}, the sound velocity and specific heat are
compared for the two couplings, as in Fig.~\ref{Fig2}. The sound
velocity reflects the stiffness of the equation of state, or
determines the flow properties in heavy-ion reactions. In the left
panel of Fig.~\ref{Fig3}, the sound velocity increases and gradually
approaches the relativistic limit $c_s^2=1/3$ as the temperature
increases. In the temperature range  $0.4\sim 0.8 T_0$, we can see
that the dashed line for the running coupling is always above the
solid line for the fixed-coupling case. In fact, at high
temperature, the quark mass is much less in the running-coupling
case than in the fixed-coupling case. The quarks with very small
masses and weak interaction strengths display a behavior similar to
the massless particles. So the running coupling induces a faster
approach to the relativistic limit at lower temperature. In the
right panel, the specific heat is shown as a function of
temperature, where the ratio of the specific heat density $C_V$ and
the cubic temperature $T^3$ is introduced as a dimensionless
quantity. The nonmonotonic shape of $C_V/T^3$ appears in both
coupling cases. But for the running coupling, the position of the
maximum of the specific heat moves in the direction of lower
temperature, which would signify that the crossover temperature may
decrease in the running-coupling case compared to the fixed-coupling
case. At very high temperature, the two lines coincide and the
specific heat maintains an almost constant value of $C_V/T^3\approx
28$, which indicates an equilibrium state of thermal radiation.

\subsection{In a strong magnetic field}

\begin{figure}[tbp]
\begin{center}
\includegraphics[width=0.40 \textwidth]{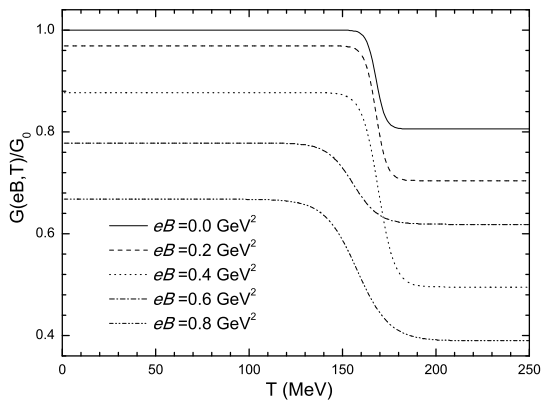}
\end{center}
\caption{{\protect\footnotesize The reduced running coupling constant $%
G(eB,T)/G_0$ as a monotonous decreasing function of the
temperature for several fixed magnetic fields.}} \label{Fig4}
\end{figure}

It is well known that the dynamical quark mass and vacuum structure
are drastically changed by a strong magnetic field and many
interesting properties are revealed. In particular, the
pseudocritical temperature for the chiral restoration transition
characterized by the inverse magnetic catalysis is a hot topic.
Inverse magnetic catalysis can be interpreted by a
magnetic-field-dependent coupling. In Fig.~\ref{Fig4}, we plot the
reduced coupling $G(eB, T)/G_0$ versus the
temperature at different magnetic fields $eB=0$, 0.2, 0.4, 0.6, and $0.8$~GeV$%
^2$. The coupling constant remains invariant when the temperature is
smaller than $140$ MeV. Moreover, the stronger the magnetic field,
the smaller the coupling constant. Then, there is a sharp drop on
each line at a critical temperature in the range of ($150\sim 170$
MeV), which is essentially determined by the parameter $T_a$ in the
coupling constant (\ref{eq:Grun}). Due to the nonmonotonous
parameter set of $T_a$ in the coupling constant (\ref{eq:Grun}), the
two lines for B=$0.4$ and $0.6$ GeV$^2$ cross each other. As in Ref.
\cite{Farias16}, the quark dynamical mass and the condensate
decrease continuously when the temperature increases. So far, a
large number of lattice simulations have demonstrated that there is
only an energy density jump (and not a true phase transition) when
the baryon chemical potential vanishes. This signals a crossover
characterized by a pseudocritical temperature $T_\mathrm{pc}$ which
is about $160$ MeV with systematic errors. The effect of the
coupling constant running with the temperature was investigated and
the entropy density can be greatly increased \cite{Farias16}. In
this section, we focus on its effect on the crossover pseudocritical
transition, which is determined by the peaks in the
susceptibilities. In the following, we define two criteria to
calculate the pseudocritical transition temperatures£¬

\begin{enumerate}
\item Criterion I: The temperature $T_\mathrm{pc}$ at which the maximum of the
derivation of the quark condensate $\phi$ with respect to the temperature
occurs,
\begin{equation}
\frac{\partial^2 \phi}{\partial T^2}=0.  \label{eq:cri1}
\end{equation}

\item Criterion $\Pi$: The temperature $T_\mathrm{pc}$ at which the maximum of $-dM/dT$ occurs,
\begin{equation}
-\frac{d^2M}{dT^2}= 2 G\frac{\partial^2 \phi}{\partial T^2}+2 \phi \frac{%
\partial^2 G}{\partial T^2} +4 \frac{\partial G}{\partial T}\frac{\partial
\phi}{\partial T}=0.  \label{eq:cri2}
\end{equation}
\end{enumerate}
Because the contribution of the last two terms in Eq.
(\ref{eq:cri2}) cannot be neglected numerically. So the two criteria
in Eqs.(\ref{eq:cri1}) and (\ref{eq:cri2}) can not be satisfied
simultaneously. Even for the coupling constant $G(B,T)=G(B)(1-\gamma
T |eB|/\Lambda_{QCD}^3)$ \cite{Farias}, the second term is zero, but
the third term will not vanish yet.

\begin{figure}[tbp]
\begin{center}
\includegraphics[width=0.40 \textwidth]{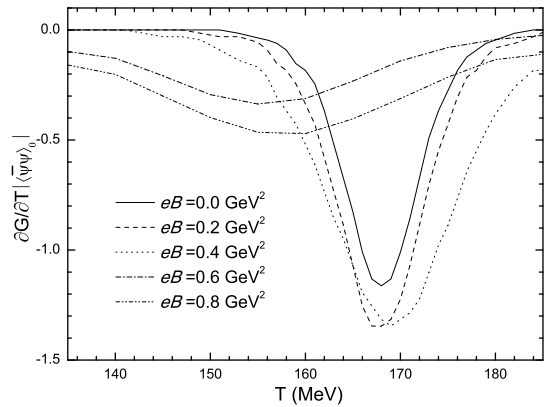}
\end{center}
\caption{{\protect\footnotesize The derivative of the coupling
constant with respect to the temperature, which is multiplied by the
vacuum quark condensate $|\langle
\bar{\protect\psi}\protect\psi\rangle_0|$ to give a dimensional
quantity.}} \label{Fig5}
\end{figure}

In Fig.~\ref{Fig5} the derivation of the coupling constant $G(eB,T)$
with temperature is shown. In order to get a dimensional quantity,
the quark condensate value in vacuum $|\langle
\bar{\psi}\psi\rangle_0|=(236.4$~MeV$)^3$ is multiplied in the
production. From the numerical result, it is obvious that the
minimum value of the derivative $\partial G/\partial T$ occurs in
the temperature range 140- 175 MeV, which always covers the range of
the pseudocritical temperature. Consequently, it is inevitable that
the derivative term $\partial G/\partial T$ will affect the position
of the crossover pseudocritical temperature $T_\mathrm{pc}$. The
parametrization of the temperature- and magnetic-field-dependent
function from Ref. \cite{Farias} will lead to a constant value for
the second term in Eq. (\ref{eq:new}). Other attempts in the
literature with a temperature-dependent coupling have also led to
considerable changes \cite{bern87}. Because $T_a$ is nonmonotonous
in Eq.~(\ref{eq:Grun}), the behavior of the curves as the magnetic
field increases is not regular.

\begin{figure}[tbp]
\begin{center}
\includegraphics[width=0.42 \textwidth]{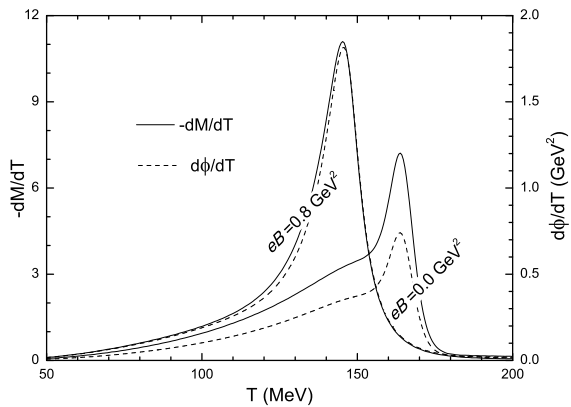}
\end{center}
\caption{{\protect\footnotesize The pseudocritical temperature
determined by the peaks of the derivatives -$dM/dT$ and
$d\protect\phi/dT$ is shown.}} \label{Fig6}
\end{figure}

Since the behavior of the coupling with the temperature cannot be
neglected, we compare the two criteria $dM/dT$ and $d\phi/dT$ in the
calculation of the pseudocritical temperature in Figs. \ref{Fig6} and \ref%
{Fig7}. First, we show the contribution of the running
coupling constant to the effective susceptibilities at the two strengths $eB=0$ and $%
eB=0.8$ GeV$^2$ in Fig.~\ref{Fig6}. For convenience of comparison,
the negative derivative $-dM/dT$ (which is dimensionless) is plotted
on the left axis, while the criterion $d\phi/dT$ (in units of
GeV$^2$) is plotted on the right axis. The peaks of the
susceptibility based on the two criteria are no longer exactly
coincident, which is particularly noticeable for the magnetic field
$eB=0.8$ GeV$^2$.

\begin{figure}[tbp]
\begin{center}
\includegraphics[width=0.40 \textwidth]{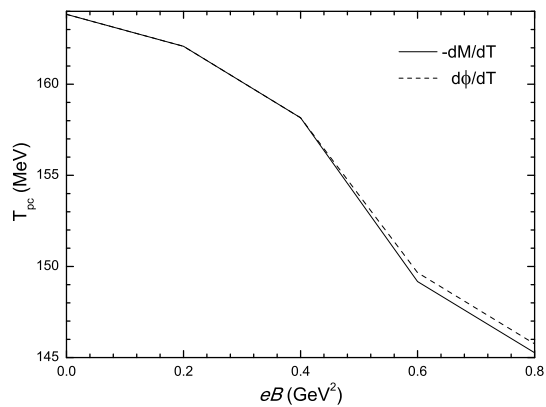}
\end{center}
\caption{{\protect\footnotesize The pseudocritical temperature
determined by the peaks of the derivatives -$dM/dT$ and
$d\protect\phi/dT$.}} \label{Fig7}
\end{figure}

Inverse magnetic catalysis can be explained by the dependence of the
QCD coupling on the strong magnetic field. At finite temperature and
vanishing density, it is further understood from the temperature-
and magnetic-field-dependent coupling that the pseudocritical
temperature decreases as the magnetic field increases. In order to
account for the effect of $G(eB,T)$ and display the difference
between the two criteria, we show the descending lines of the pseudocritical temperature $%
T_\mathrm{pc}$ as the magnetic field increases in Fig.~\ref{Fig7}.
The solid and dashed lines are derived from the peaks of the
derivatives $-dM/dT $ and $d\phi/dT$, respectively, in
Fig.~\ref{Fig6}. For weak magnetic fields, the tiny difference can
be neglected. As the magnetic field increases, the two lines are
distinctly separated. It can be clearly seen at stronger magnetic
fields that the criterion $dM/dT$ will give a lower pseudocritical
temperature $T_\mathrm{pc}$ and the inverse catalysis effect becomes
more prominent. The parametrization of the running coupling is
derived from the lattice simulation of the QCD phase diagram, and in
turn it will influence the QCD pseudocritical temperature. However,
the difference between the two criteria is less than the lattice
error. One cannot make a conclusion about which criterion is better
to get the pseudocritical temperature. To some extent, the relation
can be used to check the discrepancy between two methods for the
running coupling proposed in future work.

\section{summary}

In this paper we have employed the SU(2) NJL model to study QCD
matter with a temperature- and/or magnetic-field-dependent coupling.
Compared to the fixed-coupling constant, the temperature-dependent
coupling drastically changes the dynamical quark mass, entropy
density, sound velocity, and specific heat. As the temperature
increases up to $T_0$, the entropy density, sound velocity, and
specific heat density go up and approach to the critical values
$S/T^3\approx9.2$, $c_s=\sqrt{1/3}$ and $C_V/T^3\approx 28$. In the
temperature range $0.4\sim 0.8 T_0$, we found that the entropy and
sound velocity in the running-coupling case are remarkably larger
than those in the fixed-coupling case.

It is also helpful to use the magnetic-field- and
temperature-dependent coupling when accounting for inverse magnetic
catalysis at finite temperature. In previous work, the position of
the crossover transition characterized by a pseudocritical
temperature was determined by the peak of the susceptibility of the
quark condensate or the quark dynamical mass with respect to the
temperature. The two criteria $d\phi/dT$ and $dM/dT$ are coincident
when the coupling is independent of the temperature. However, when
the coupling constant depends on the temperature and magnetic field,
a discrepancy will appear between the two criteria due to the
presence of the nonzero term $\partial G(eB,T)/\partial T$. The
criterion $dM/dT$ leads to a lower pseudocritical temperature for
the crossover transition. The special value of $T_\mathrm{pc}$ will
depend on the running behavior of the coupling constant with the
temperature. Therefore, we argued that possible choices for the
coupling constant in future work will influence the QCD phase
diagram in turn.

Up to now, spontaneous chiral symmetry breaking has been studied in
backgrounds of electric and magnetic field \cite{rugg16}. The QCD
phase diagram could be ruled by the rather complicated interaction
structure. A careful treatment should be done in the case of the
temperature- and magnetic-field-dependent coupling $G(eB,T)$. In a
future paper, we hope to consider the implications for the color
superconducting phase to develop a deeper understanding of the QCD
phase diagram.

\begin{acknowledgments}
The authors would like to thank support from the National Natural
Science Foundation of China under the Grant Nos. 11475110, 11575190,
and 11705163. This work was also sponsored by the Fund for Shanxi
¡°1331 Project¡± Key Subjects Construction.
\end{acknowledgments}

\end{document}